\documentclass[twocolumn,showpacs,prx,aps,longbibliography,superscriptaddress,nofootinbib]{revtex4-2}
\usepackage[bookmarks=false,linkcolor=blue,urlcolor=blue,colorlinks,citecolor=blue]{hyperref}
\usepackage{array}
\usepackage{tabularray}
\pdfoutput=1
\usepackage{xcolor}
\usepackage{placeins}
\usepackage{svg}
\usepackage{adjustbox}
\usepackage{graphicx}
\usepackage{color}
\usepackage{amsmath}
\usepackage{amsfonts}
\usepackage{mathrsfs}
\usepackage{amsthm}
\usepackage{amssymb}
\usepackage{wrapfig}
\usepackage[shortlabels]{enumitem}
\usepackage[retainorgcmds]{IEEEtrantools}
\usepackage{tcolorbox}
\usepackage{enumitem}
\usepackage{booktabs}
\usepackage{algpseudocode}

\makeatletter
\AtBeginDocument{\let\LS@rot\@undefined}
\makeatother
\newcolumntype{C}[1]{>{\centering\arraybackslash}m{#1}}
\begin{document}
	\author{Lior Oppenheim}
	\affiliation{The Racah Institute of Physics, The Hebrew University of Jerusalem, Jerusalem 9190401, Israel}
	\author{Snir Gazit}
\affiliation{The Racah Institute of Physics, The Hebrew University of Jerusalem, Jerusalem 9190401, Israel}
\affiliation{The Fritz Haber Research Center for Molecular Dynamics, The Hebrew University of Jerusalem, Jerusalem 9190401, Israel}
	\author{Zohar Ringel}
\affiliation{The Racah Institute of Physics, The Hebrew University of Jerusalem, Jerusalem 9190401, Israel}
        \title{Improving CFT Operators Using Machine Learning}
	
\begin{abstract}
Finite-size effects limit the accuracy with which conformal data can be extracted from lattice simulations of critical systems. While action improvement suppresses some corrections to scaling, it does not address operator-dependent effects arising from imperfect lattice representations of continuum conformal fields. In this work, we propose a data-driven method for improving lattice operators themselves, constructing estimators with enhanced overlap with the corresponding primary operators of the continuum conformal field theory. We identify improved lattice representations of leading spin and energy operators in three two-dimensional critical systems: the Ising model, the $q=3$ Potts model, and the dilute $q=3$ Potts model. In all cases, the resulting operators exhibit reduced corrections to scaling and yield more accurate estimates of scaling dimensions compared to conventional lattice choices. The code and analysis workflows used to produce these results are made available in an accompanying GitHub repository.
\end{abstract}
	
\maketitle
 
\textit{Introduction - }  Bridging the gap between lattice formulations of quantum and statistical systems and their continuum field-theory descriptions remains a central challenge in modern physics. Numerical studies, whether performed on classical or quantum computers, are inherently constrained by finite system sizes and limited computational resources. Consequently, attempts to extract infrared (IR) observables from ultraviolet (UV) simulations are inevitably affected by cutoff or discretization effects, which manifest as systematic deviations from continuum behavior \cite{navas_2024}. In critical systems, these appear as finite-size corrections to scaling, typically governed by power-law forms \cite{cardy_1988}. Such corrections can be severe: in some cases, they obscure the values of critical exponents and even the nature of the phase transition itself, as observed in studies of deconfined quantum criticality \cite{nahum_2015}.

Progress, therefore, hinges on strategies that accelerate convergence to IR physics beyond the naive approach of simply simulating ever-larger lattices. A well-established route to action improvement is fine-tuning additional symmetry-allowed couplings, which effectively cancel specific irrelevant perturbations and thus suppresses the leading corrections to scaling \cite{hasenfratz_1993,campostrini_1999,hasenbusch_1999,segall_2025,hasenbusch_2025}. From a renormalization-group (RG) perspective, this corresponds to tuning the couplings of irrelevant operators to zero, thereby moving the critical point closer to the continuum fixed point. Although successful in many contexts, this strategy has clear limitations: it requires identifying and tuning the relevant couplings on a case-by-case basis.

In this work, we pursue a complementary approach - improving the operators themselves \cite{luscher_1998,heatlie_1991,sandvik_2025}. For a fixed lattice action, we construct estimators that maximize overlap with the true conformal operators of the continuum theory. Concretely, we show how to extract scaling dimensions more accurately by identifying lattice operators whose correlations are dominated by the corresponding CFT (Conformal Field Theory) primaries, thereby reducing corrections that cannot be eliminated by action improvement alone.

Our method builds on recent developments at the intersection of statistical physics and machine learning. In particular, we employ the Real-Space Mutual Information Neural Estimator (RSMI-NE) \cite{ringel_2018}, a neural-network algorithm grounded in the correspondence between information-theoretic relevance and RG relevance \cite{gordon_2021}. Under well-defined assumptions, RSMI-NE isolates the dominant eigenvector of the transfer matrix, corresponding to the leading primary in the CFT spectrum. This approach has already been used to qualitatively identify low-lying relevant operators and symmetries in systems where these were not easily accessible, for example, due to gauge symmetry \cite{oppenheim_2024}. Here we show that RSMI-NE can also make quantitative progress by detecting improved operators, which reduce scaling corrections and yield more accurate estimates of critical exponents\footnote{The code of the RSMI-NE algorithm used to generate the results of this paper is publicly available at \url{https://github.com/oppenheim/rsmi-improved-operators}.}.

\textit{Origin of corrections to scaling —}
At criticality, a two-point correlator in the continuum decays as
\[
\langle \mathcal{O}_{\mathrm{CFT}}(r_1)
\mathcal{O}_{\mathrm{CFT}}(r_2)\rangle
\propto |r_1-r_2|^{-2\Delta_{\mathcal O}},
\]
where $\Delta_{\mathcal O}$ is the scaling dimension of the primary operator.
On the lattice, this power law acquires corrections,
\begin{equation}
\label{eq:correlation}
\resizebox{\columnwidth}{!}{$
\langle \mathcal{O}_{\mathrm{lat}}(r_1)
\mathcal{O}_{\mathrm{lat}}(r_2)\rangle
\propto |r_1-r_2|^{-2\Delta_{\mathcal O}}
\left(
1 + \sum_{i,n\in\mathbb{N}} a_{i,n} |r_1 - r_2|^{-n\omega_i}+\dots
\right)$}
\end{equation}

where $\omega_i>0$ are correction-to-scaling exponents, $n$ labels the
order of the correction, and $a_{i,n}$ determines the corresponding
amplitude.

These corrections arise from two main sources. Within the Symanzik framework \cite{symanzik_1983,ramos_2016}, the lattice action and operator admit expansions:
\begin{equation*}
\resizebox{\columnwidth}{!}{$
S_{\mathrm{lat}}=S_{\mathrm{CFT}}+\sum_i \lambda_i \int d^dx\, U_i(x),
\qquad
\mathcal{O}_{\mathrm{lat}}
=\alpha_0 \mathcal{O}_{\mathrm{CFT}}
+\sum_i \alpha_i D_i,
$}
\end{equation*}
where $U_i$ are irrelevant scalar operators and $D_i$ are operators in the same symmetry sector as $\mathcal O_{\mathrm{CFT}}$.

Irrelevant operators in the action generate corrections controlled by $\omega_i=\Delta_{U_i}-d$, which are universal and independent of the specific lattice representation of $\mathcal O_{\mathrm{CFT}}$. In contrast, mixing of $\mathcal O_{\mathrm{CFT}}$ with its descendants (e.g.\ $\nabla^2 \mathcal O_{\mathrm{CFT}}$) produces \emph{analytic} \cite{cardy_1986} corrections with integer exponents, whose amplitudes depend on the coefficients $\alpha_i$. More generally, analytic corrections need not arise solely from operator mixing; irrelevant descendants in the action can also contribute integer-power terms (e.g via the $T\bar{T}$ operator). Thus, an appropriate choice of lattice operator can reduce the operator-dependent analytic corrections by minimizing descendant mixing.

\textit{Statement of the Problem - } 
We consider a Monte Carlo simulation of lattice models at criticality, and focus on the estimation of scaling dimensions of the corresponding 2d CFT's primary operators $\mathcal{O}_{CFT}$. On the lattice, these are accessed through two-point correlation functions of local operators $\mathcal{O}_{lat}$ measured under the lattice action $S_{lat}$. Specifically, we consider two-point connected correlators of the form of eq.~\ref{eq:correlation}. By fitting the data, one can, in principle, extract both $\Delta_{\mathcal{O}}$ and $\omega_i$.

Yet the choice of the lattice operator is far from unique, and different realizations of the same continuum operator can exhibit different finite-size effects. This raises the central question: Can one identify an improved lattice representation of $\mathcal{O}_{CFT}$ that suppresses the leading analytical corrections to scaling coefficients and thus yields cleaner access to its scaling dimension?

\textit{Models -} For this purpose, we will study improved representations of lattice operators in three two-dimensional statistical systems on a square lattice: Ising, $q=3 $ Potts, and dilute $q=3$ Potts models. These models were chosen for their simplicity and for the a priori knowledge of their CFT theory (i.e., scaling dimensions of all operators).

The \textit{Ising} model consists of the following action ($\sigma_i\in\{-1,1\}$):
$$-\beta H_{\text{ising}}=K\sum_{\left\langle i,j\right\rangle }\sigma_{i}\sigma_{j}$$
which reaches criticality at $K_c=\frac{1}{2}\log\left(1+\sqrt{2}\right)$. The two leading primary operators for the 2d Ising critical theory are: $\sigma$ (spin operator) and $\epsilon$ (energy operator), with corresponding conformal dimensions of $\frac{1}{8}$ and $1.0$. Since there are no primary irrelevant operators, all corrections to scaling exponents are analytic with $\omega=2,4,\dots$.

The $q=3$ \textit{Potts} model consists of the following action ($\sigma_i\in\{1,2,3\}$):
$$-\beta H_{\text{q=3 Potts}}=K\sum_{\left\langle i,j\right\rangle}\limits\delta_{\sigma_i,\sigma_j}$$
which reaches criticality at $K_c=\log\left(1+\sqrt{3}\right)$.  The two leading primary operators for the $q=3$ 2d Potts critical theory are: $\sigma$ and $\epsilon$, with corresponding conformal dimensions of $\frac{2}{15}$ and $\frac{4}{5}$. The corrections to scaling exponents include a non-analytic exponent $\omega=4/5$ from the leading irrelevant operator, and analytic corrections with $\omega = 2,4,\dots$.  

The \textit{dilute Potts} model is a variant of the standard $q=3$ Potts model, obtained by adding a site-dilution (chemical-potential) term to the Hamiltonian:
$$-\beta H=K\sum_{\left<i,j\right>}\limits\delta_{\sigma_i,\sigma_j}\left(1-\delta_{\sigma_i,0}\right)+D\sum_i\limits\delta_{\sigma_i,0}$$
which reaches criticality at $K_c=1.16941(2),D_c=1.37655(5)$ \cite{qian_2005}. At this point, the amplitude of the leading irrelevant operator vanishes, eliminating the dominant $\omega=0.8$ correction to scaling presented in the regular Potts model \cite{qian_2016}. Otherwise, the operator spectrum is identical to that of a regular $q=3$ Potts model.

\begin{figure*}[t]
    \centering
\hrule
    \vspace{0.4pt}
    \hrule

    \vspace{1.0ex}
    \adjustbox{trim={0.00\width} {0.00\height} {0.00\width} {0.10\height},clip}{\includegraphics[width=\textwidth]{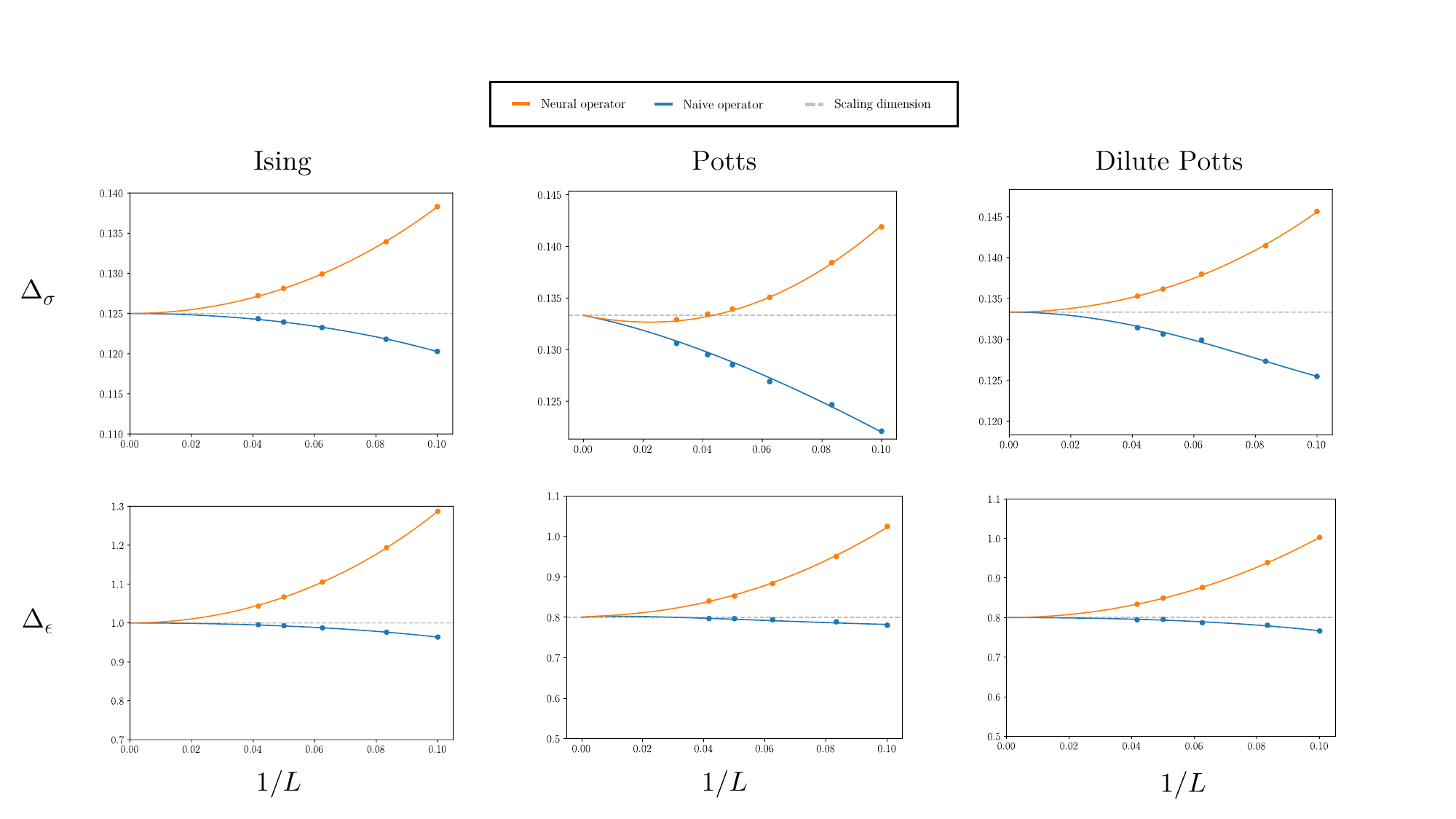}}

    \vspace{1.0ex}
    \hrule
    \vspace{0.4pt}
    \hrule

    \caption{Finite-size scaling of two-point correlation functions using Sandvik's method (see eq.~\ref{sandvik_scaling}). The six panels show the scaling for the leading spin ($\sigma$) and energy ($\epsilon$) operators in the 2d Ising, $q=3$ Potts, and dilute $q=3$ Potts models, comparing naive lattice operators with machine-learned (neural) estimators.
For the Ising and dilute Potts models, the data are fitted to $a_1 L^{-2} + a_2 L^{-4} + \Delta_{\mathcal O}$, reflecting purely analytic corrections to scaling. For the $q=3$ Potts model, the fit includes both non-analytic and analytic terms, $a_1 L^{-0.8} + a_2 L^{-1.6} + a_3 L^{-2} + \Delta_{\mathcal O}$. The coefficients of the non-analytic corrections ($L^{-0.8}$ and $L^{-1.6}$) are constrained to be identical for the naive and neural operators, as these originate from irrelevant operators in the action and are, to a leading order, independent of the operator definition. The leading analytic correction coefficient ($L^{-2}$) is strongly suppressed for the neural operators across all systems. Expressed as the ratio of absolute coefficients (neural/naive), the reduction is $\sim 0.12$ for the Ising energy operator and $\sim 0.33$ for the Ising spin operator.
For the $q=3$ Potts model, the corresponding ratios are $\sim 0.23$ for the energy operator and $\sim 0.10$ for the spin operator.
In the dilute Potts model, the ratios are $\sim 0.10$ for the energy operator and $\sim 0.91$ for the spin operator.}
    \label{fig:scaling_panels}
\end{figure*}

\textit{Methods - } We focus on the two leading primary operators in each theory. To find an improved representation, we employ RSMI-NE, a deep neural network algorithm with theoretical guarantees \cite{gordon_2021}, which produces operators ordered by their RG relevance. The resulting “neural” operator (i.e., a deep neural network acting on the degrees-of-freedom and outputting the operator value) can then be compared with the conventional (naive) lattice choice (e.g.\ the average magnetization for the spin operator). Each neural operator acts on a circular neighborhood of radius $r=3$, i.e., on all lattice degrees of freedom within distance $r$ from the operator center (37 spins in total for $r=3$). To prevent symmetry-induced mixing between operators, we project the learned operators onto their respective symmetry sectors. For instance, the rotationally invariant $\mathbb{Z}_2$-odd subspace for the Ising spin and the rotationally invariant $\mathbb{Z}_2$-even subspace for the Ising energy operator. More details regarding the exact RSMI-NE algorithm implementation appear in Appendix \ref{sec:rsmi_details}.

To quantify the improvement, we compare the scaling behavior of the neural operator against that of the naive lattice operator. To that extent, we employ Sandvik’s finite-size scaling method \cite{sandvik_2020}. For two systems of linear sizes $L$ and $2L$ and an operator $O$,one can approximate the scaling dimension of it $\Delta_O$ using this formula (derived from eq.~\ref{eq:correlation}): 
\begin{equation}
\label{sandvik_scaling}
\frac{1}{2}\log_{2}\left(\frac{\left\langle O\left(0,0\right)O\left(\frac{L}{2},\frac{L}{2}\right)\right\rangle _{L}}{\left\langle O\left(0,0\right)O\left(\frac{2L}{2},\frac{2L}{2}\right)\right\rangle _{2L}}\right)=\Delta_{O}+\sum_{n=1}^{\infty}\frac{a_{n}}{L^{\omega_{n}}}    
\end{equation}
Where the $\omega_n$'s are the corrections to scaling exponents and the $a_n$ are their magnitudes.

\begin{table}[t]
    \centering
    \begin{tabular}{llcc}
        \hline
        Operator & Analytical value & Neural operator & Naive operator \\
        \hline
        \multicolumn{4}{l}{\textbf{Ising}} \\
        $\epsilon$ & $1$ & $\boldsymbol{1.0 \pm 0.001}$ & $1.008 \pm 0.005$ \\
        $\sigma$   & $1/8 = 0.125$ & $\boldsymbol{0.1251 \pm 0.0003}$ & $0.1255 \pm 0.0003$ \\
        \hline
        \multicolumn{4}{l}{\textbf{Potts}} \\
        $\epsilon$ & $4/5 = 0.8$ & $\boldsymbol{0.797 \pm 0.002}$ & $0.838 \pm 0.003$ \\
        $\sigma$   & $2/15 = 0.133\dots$ & $\boldsymbol{0.133 \pm 0.001}$ & $0.1324 \pm 0.0002$ \\
        \hline
        \multicolumn{4}{l}{\textbf{Dilute Potts}} \\
        $\epsilon$ & $4/5 = 0.8$ & $\boldsymbol{0.798 \pm 0.004}$ & $0.79 \pm 0.01$ \\
        $\sigma$   & $2/15 = 0.133\dots$ & $\boldsymbol{0.1329 \pm 0.0009}$ & $0.1339 \pm 0.0006$ \\
        \hline
    \end{tabular}
    \caption{ Comparison of scaling-dimension estimates obtained from finite-size fits of the form $a L^{-\omega} + \Delta_{\mathcal O}$, where $a$, $\omega$, and $\Delta_{\mathcal O}$ are treated as free parameters. For each system and operator, results are shown for both the neural estimator and the conventional (or naive) lattice operator. Boldface indicates the estimator with the smaller root-mean-square (RMS) deviation from the analytical value, computed assuming a Gaussian distribution for the quoted uncertainty.}
    \label{tab:scaling_table}
\end{table}

\begin{figure}
    \centering
    \adjustbox{trim={0.03\width} {0.15\height} {0.03\width} {0.20\height},clip}{
        \includegraphics[width=0.5\textwidth]{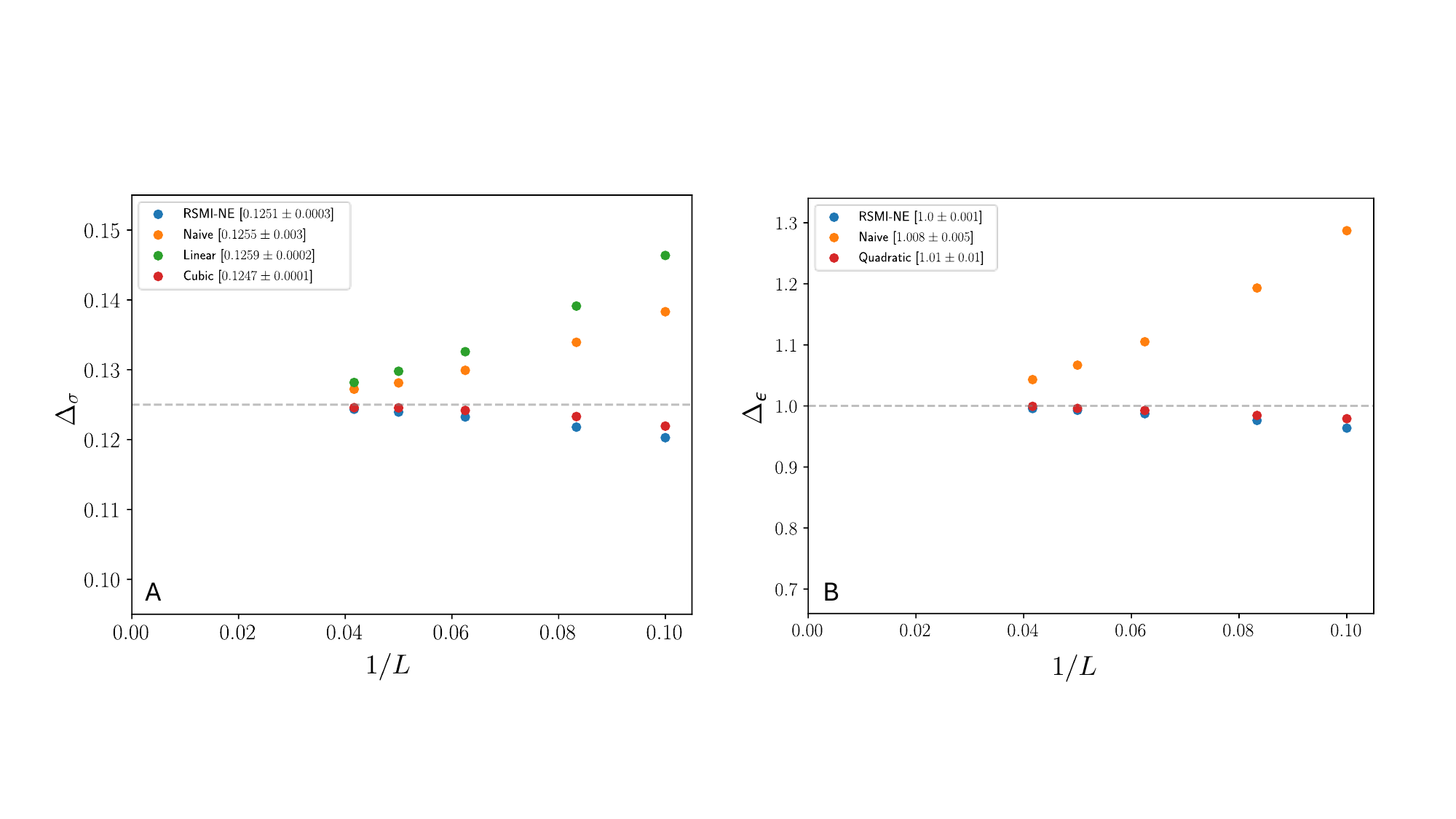}
    }
    \caption{Scaling performance of low-order surrogates for the learned Ising operators. (A) Finite-size scaling analysis for the spin operator $\sigma$, comparing the naive lattice estimator, the neural estimator, and polynomial surrogates obtained by fitting the neural operator to linear and linear plus cubic forms. The linear surrogate's scaling resembles that of the naive estimator, indicating that cubic terms are required to reproduce the neural operator's improved scaling behavior.  (B) Corresponding analysis for the energy operator $\epsilon$, comparing the naive estimator, the neural estimator, and a quadratic surrogate. The quadratic surrogate closely matches the neural estimator, indicating that quadratic terms already capture the learned improvement. In the legend, brackets indicate the scaling dimensions extracted from the finite-size fits ($aL^{-\omega}+\Delta_\mathcal{O}$).}
    \label{fig:graphs}
\end{figure}

\textit{Results -} We evaluate the quality of the neural operator in comparison to the naive approach using several benchmarks. In Figure \ref{fig:scaling_panels}, we fit the amplitudes of the corrections to determine the scaling exponent for three different systems, while keeping both the scaling dimension and the corrections to scaling exponents fixed at their known values. In the Potts model, the $\omega=0.8,1.6$ terms are taken to be action-induced and therefore common to the two operators at leading order, while the $\omega=2.0$ term is allowed to differ. We observe a consistent reduction in the amplitude of the leading analytical correction. Notably, this effect is more pronounced for the energy operator ($\epsilon$) than for the spin operator ($\sigma$), a point that will be addressed in the discussion section.

In a second benchmark, we fit the data to the general form $aL^{-\omega}+\Delta_\mathcal{O}$ -  Estimating the scaling dimension without assuming any prior knowledge of the scaling dimension nor the leading corrections to the scaling exponent, akin to more realistic scenarios. In this case, we also see a significant improvement of the neural operator compared to the naive one, especially with respect to the $\epsilon$ operator. In all cases, the neural operator yields a better estimation than the naive one: Either both hit the correct scaling dimension and the neural one has a smaller error, or only the neural estimator hits the correct scaling dimension within its margins of error.

In a third benchmark, we compare the correlation decay between the two operators for a given system size, where both operators' variances are normalized. For the Ising case, we see a steady ratio of $\times 3.5$ between the two correlations at long ranges for the $\epsilon$ operator and $\times 1.1$ for the $\sigma$ operator. In both the Potts and the dilute Potts cases, we see a ratio of $\times 2.5$ for the $\epsilon$ operator and $\times 1.05$ for the $\sigma$ operator. See Appendix \ref{sec:correlation_comparison} for additional details. These results correlate well with the expected output of the RSMI-NE algorithm: In leading order, the stronger the long-range correlation, the more information the operator can retain from its environment.

\textit{Ising Case study} -  Here, we explore which features are beneficial for improved scaling dimension. Our strategy is to construct explicit low-order surrogates for the RSMI-NE operators learned in the Ising critical theory. Concretely, we fit each neural operator to a symmetry-allowed polynomial in the local degrees of freedom within the spin field. Symmetry restricts the expansion: in the Ising $\mathbb{Z}_2$-odd sector (e.g. the spin operator $\sigma$) only odd powers of the spin field are allowed, and in the $\mathbb{Z}_2$-even sector (e.g. the energy operator $\epsilon$), only even powers.

We first assess how well these low-order surrogates reproduce the scaling behavior of the neural estimators. Figure~\ref{fig:graphs}A shows that, for $\sigma$, the linear surrogate behaves similarly to the naive lattice operator and does not capture the neural estimator's improved finite-size scaling, whilst adding the leading $\mathbb{Z}_2$-odd nonlinear correction (of ~50,000 possible cubic terms) brings the surrogate into agreement with the neural scaling trend. In contrast, Fig.~\ref{fig:graphs}B shows that, for $\epsilon$, a quadratic surrogate (of ~1,000 parameters) largely reproduces the neural estimator's scaling behavior. Moreover, in all cases, the surrogates' scaling dimension estimation is worse than that of the full operator, demonstrating the necessity of including higher-order polynomials to reach optimal results.

\begin{figure}[t]
    \centering
    \adjustbox{trim={0.04\width} {0.20\height} {0.03\width} {0.20\height},clip}{
        \includegraphics[width=0.50\textwidth]{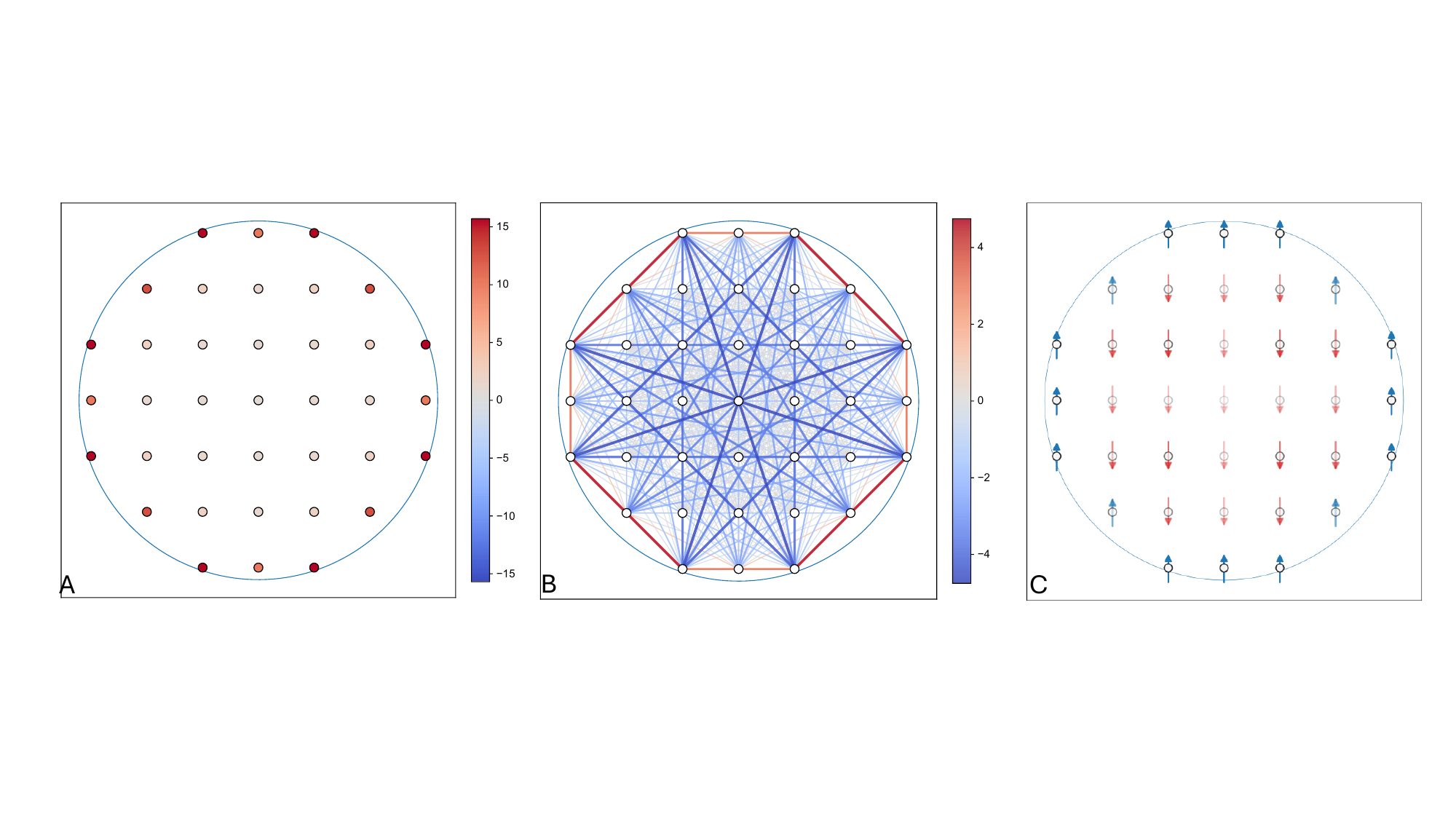}
    }
    \caption{Low-order structure of the fitted surrogates within the spin field. 
    (A) Linear filter for the $\mathbb{Z}_2$-odd spin operator $\sigma$, shown as site weights. 
    (B) Quadratic surrogate for the $\mathbb{Z}_2$-even energy operator $\epsilon$, visualized as a weighted graph over pairs of sites (edge color encodes sign and magnitude).
    (C) A representative spin configuration that maximizes the cubic features of the $\sigma$ surrogate; the minimizing configuration follows from a global $\mathbb{Z}_2$ flip. Spin transparency encodes local sensitivity: more opaque sites are those whose flip would produce a larger decrease in the cubic component (i.e., they contribute more strongly to the maximum).}
    \label{fig:filters}
\end{figure}

To connect these scaling observations to real-space structure, we visualize the fitted coefficients in Fig.~\ref{fig:filters}. The linear surrogate for $\sigma$ corresponds to a real-space filter over the spin field (Fig.~\ref{fig:filters}A). The quadratic surrogate for $\epsilon$ is naturally represented as a pairwise coupling graph (Fig.~\ref{fig:filters}B). Finally, to build intuition for the leading nonlinear correction in the $\sigma$ sector, we display in Fig.~\ref{fig:filters}C a configuration that extremizes just the cubic part of the surrogate (the opposite extremum follows from the global $\mathbb{Z}_2$ symmetry). 

A recurring feature across panels is that the dominant weights/couplings are concentrated at the boundary of the operator, consistent with \cite{gordon_2021}. For the cubic case, we get that in the extremal configuration, the spins at the boundary have an opposite sign to that of the bulk. Since averaging $\nabla^2\sigma$ over the spin field is maximized exactly by such a configuration (due to Gauss’s theorem), we hypothesize that this component reduces the descendant $\nabla^2\sigma$ component of the CFT operator.

\textit{Discussion} -In this work, we have shown that RSMI-NE can be used to construct improved lattice representations of CFT primary operators. In the Ising, $q=3$ Potts, and dilute $q=3$ Potts models, the learned operators consistently reduce analytic corrections to scaling relative to conventional lattice definitions. The effect is especially pronounced for the energy operator $\epsilon$, while improvements for the spin operator $\sigma$ are more modest. 

Within the Symanzik framework, this improvement can be interpreted as a suppression of descendant mixing in the lattice operator expansion. The learned operators appear to reduce the amplitude of these terms, effectively increasing overlap with the CFT primary. The greater improvement for $\epsilon$ suggests that the naive lattice energy operator contains larger descendant admixture than the magnetization does for $\sigma$. A quantitative field-theoretic understanding of this asymmetry remains open.

A recurring structural feature of the learned operators is the concentration of weight near the boundary of their support.  This boundary localisation was rigorously derived only for infinite-cylinder geometries in Ref.~\cite{gordon_2021}. Our results show that it persists on finite planar lattices. This robustness may reflect the conformal equivalence between cylinder and plane geometries via $z = e^{w}$, suggesting that the information-theoretic criterion used by RSMI-NE captures intrinsic CFT structure rather than geometry-specific artifacts.

Finally, although low-order polynomial surrogates reproduce part of the neural operators’ behavior, they require thousands of parameters to approach the neural scaling performance. This indicates that the improvement is not attributable to a simple low-dimensional Ansatz, but rather to a structured combination of many symmetry-allowed contributions that would be difficult to engineer manually. In this sense, the neural network serves as a variational optimizer within a very high-dimensional operator space.

\textit{Outlook} - This work continues the exploration of the RSMI-NE algorithm, which has shown promising results since its inception \cite{ringel_2018,efe_2021,efe2_2021,efe_2023,oppenheim_2024}. A natural next step is to apply it in higher-dimensional systems where scaling dimensions are known only numerically. In these more numerically demanding scenarios, operator improvement may advance our current precision. 
Furthermore, Operator and action improvement address complementary sources of corrections to scaling, suggesting a combined approach for precision studies of exotic critical phenomena. Turning to qualitative aspects, it would be interesting to consider extensions of the RSMI-NE for detecting non-local observables (e.g., string operators) and operators in fermionic systems (e.g. \cite{Tong_2022}) where the operator content remains to be mapped out. 

\FloatBarrier

\bibliography{bibliography}

@article{Tong_2022,
   title={Comments on symmetric mass generation in 2d and 4d},
   volume={2022},
   ISSN={1029-8479},
   url={http://dx.doi.org/10.1007/JHEP07(2022)001},
   DOI={10.1007/jhep07(2022)001},
   number={7},
   journal={Journal of High Energy Physics},
   publisher={Springer Science and Business Media LLC},
   author={Tong, David},
   year={2022},
   month=jul }

@article{nahum_2015,
  title = {Deconfined Quantum Criticality, Scaling Violations, and Classical Loop Models},
  author = {Nahum, Adam and Chalker, J. T. and Serna, P. and Ortu\~no, M. and Somoza, A. M.},
  journal = {Phys. Rev. X},
  volume = {5},
  issue = {4},
  pages = {041048},
  numpages = {21},
  year = {2015},
  month = {Dec},
  publisher = {American Physical Society},
  doi = {10.1103/PhysRevX.5.041048},
  url = {https://link.aps.org/doi/10.1103/PhysRevX.5.041048}
}

@article{qian_2005,
  title = {Dilute Potts model in two dimensions},
  author = {Qian, Xiaofeng and Deng, Youjin and Bl\"ote, Henk W. J.},
  journal = {Phys. Rev. E},
  volume = {72},
  issue = {5},
  pages = {056132},
  numpages = {15},
  year = {2005},
  month = {Nov},
  publisher = {American Physical Society},
  doi = {10.1103/PhysRevE.72.056132},
  url = {https://link.aps.org/doi/10.1103/PhysRevE.72.056132}
}

@article{sandvik_2020,
    author = {Anders W. Sandvik and Bowen Zhao},
    title = {Consistent Scaling Exponents at the Deconfined Quantum-Critical Point},
    publisher = {Chin. Phys. Lett.},
    year = {2020},
    journal = {Chinese Physics Letters},
    volume = {37},
    number = {5},
    pages = {057502},
    url = {https://cpl.iphy.ac.cn/EN/abstract/article_105571.shtml}
}

@article{navas_2024,
  title = {Review of Particle Physics},
  author = {Navas, S. et. al.},
  collaboration = {Particle Data Group Collaboration},
  journal = {Phys. Rev. D},
  volume = {110},
  issue = {3},
  pages = {350-366},
  numpages = {17},
  year = {2024},
  month = {Aug},
  publisher = {American Physical Society},
  doi = {10.1103/PhysRevD.110.030001},
  url = {https://link.aps.org/doi/10.1103/PhysRevD.110.030001}

}

@article{cardy_1986,
title = {Operator content of two-dimensional conformally invariant theories},
journal = {Nuclear Physics B},
volume = {270},
pages = {186-204},
year = {1986},
issn = {0550-3213},
doi = {https://doi.org/10.1016/0550-3213(86)90552-3},
url = {https://www.sciencedirect.com/science/article/pii/0550321386905523},
author = {John L. Cardy}
}

@book{cardy_1988,
  editor       = {Cardy, John L.},
  title        = {Finite-Size Scaling},
  series       = {Current Physics - Sources and Comments},
  volume       = {2},
  publisher    = {North-Holland (Elsevier)},
  address      = {Amsterdam; Oxford; New York; Tokyo},
  year         = {1988},
  isbn         = {0-444-87109-8; 0-444-87110-1},
}

@article{campostrini_1999,
  title = {Improved high-temperature expansion and critical equation of state of three-dimensional Ising-like systems},
  author = {Campostrini, Massimo and Pelissetto, Andrea and Rossi, Paolo and Vicari, Ettore},
  journal = {Phys. Rev. E},
  volume = {60},
  issue = {4},
  pages = {3526--3563},
  numpages = {0},
  year = {1999},
  month = {Oct},
  publisher = {American Physical Society},
  doi = {10.1103/PhysRevE.60.3526},
  url = {https://link.aps.org/doi/10.1103/PhysRevE.60.3526}
}

@article{hasenbusch_1999,
  title = {Critical exponents of the three-dimensional Ising universality class from finite-size scaling with standard and improved actions},
  author = {Hasenbusch, M. and Pinn, K. and Vinti, S.},
  journal = {Phys. Rev. B},
  volume = {59},
  issue = {17},
  pages = {11471--11483},
  numpages = {0},
  year = {1999},
  month = {May},
  publisher = {American Physical Society},
  doi = {10.1103/PhysRevB.59.11471},
  url = {https://link.aps.org/doi/10.1103/PhysRevB.59.11471}
}

@article{symanzik_1983,
title = {Continuum limit and improved action in lattice theories: (I). Principles and phi 4 theory},
journal = {Nuclear Physics B},
volume = {226},
number = {1},
pages = {187-204},
year = {1983},
issn = {0550-3213},
doi = {https://doi.org/10.1016/0550-3213(83)90468-6},
url = {https://www.sciencedirect.com/science/article/pii/0550321383904686},
author = {K. Symanzik}
}

@article{oppenheim_2024,
  title = {Machine learning the operator content of the critical self-dual Ising-Higgs lattice gauge theory},
  author = {Oppenheim, Lior and Koch-Janusz, Maciej and Gazit, Snir and Ringel, Zohar},
  journal = {Phys. Rev. Res.},
  volume = {6},
  issue = {4},
  pages = {043322},
  numpages = {10},
  year = {2024},
  month = {Dec},
  publisher = {American Physical Society},
  doi = {10.1103/PhysRevResearch.6.043322},
  url = {https://link.aps.org/doi/10.1103/PhysRevResearch.6.043322}
}

@article{segall_2025,
  title = {Improved actions using the renormalization group},
  author = {Segall, Guy and Gazit, Snir and Podolsky, Daniel},
  journal = {Phys. Rev. B},
  volume = {111},
  issue = {18},
  pages = {184413},
  numpages = {18},
  year = {2025},
  month = {May},
  publisher = {American Physical Society},
  doi = {10.1103/PhysRevB.111.184413},
  url = {https://link.aps.org/doi/10.1103/PhysRevB.111.184413}
}

@article{hasenbusch_2025,
    author = {Hasenbusch, Martin},
    title = {Eliminating leading and subleading corrections to scaling in the three-dimensional XY universality class},
    year = {2025},
    journal = {arXiv:2507.19265 [cond-mat.stat-mech]},
    url = {https://doi.org/10.48550/arXiv.2507.19265}
}

@article{ringel_2018,
	year = {2018},
	publisher = {Springer Science and Business Media {LLC}},
	volume = {14},
	number = {6},
	pages = {578--582},
	author = {Maciej Koch-Janusz and Zohar Ringel},
	title = {Mutual information, neural networks and the renormalization group},
	journal = {Nature Physics},
    url = {https://www.nature.com/articles/s41567-018-0081-4}
}

@article{gordon_2021,
    publisher = {American Physical Society ({APS})},
    volume = {126},
    issue = {24},
    pages = {240601},
    numpages = {7},
    year = {2021},
    month = {Jun},
    author = {Amit Gordon and Aditya Banerjee and Maciej Koch-Janusz and Zohar Ringel},
    title = "{Relevance in the Renormalization Group and in Information Theory}",
    journal = {Phys. Rev. Lett.},
    url = {https://doi.org/10.1103/PhysRevLett.126.240601}
}

@article{qian_2016,
  title = {Equivalent-neighbor Potts models in two dimensions},
  author = {Qian, Xiaofeng and Deng, Youjin and Liu, Yuhai and Guo, Wenan and Bl\"ote, Henk W. J.},
  journal = {Phys. Rev. E},
  volume = {94},
  issue = {5},
  pages = {052103},
  numpages = {9},
  year = {2016},
  month = {Nov},
  publisher = {American Physical Society},
  doi = {10.1103/PhysRevE.94.052103},
  url = {https://link.aps.org/doi/10.1103/PhysRevE.94.052103}
}

@misc{luscher_1998,
    title={Advanced Lattice QCD}, 
    author={Martin Lüscher},
    year={1998},
    journal = {arXiv:9802029 [hep-la]},
    url = {https://doi.org/10.48550/arXiv.hep-lat/9802029}
}

@article{heatlie_1991,
title = {The improvement of hadronic matrix elements in lattice QCD},
journal = {Nuclear Physics B},
volume = {352},
number = {1},
pages = {266-288},
year = {1991},
issn = {0550-3213},
doi = {https://doi.org/10.1016/0550-3213(91)90137-M},
url = {https://www.sciencedirect.com/science/article/pii/055032139190137M},
author = {G. Heatlie and C.T. Sachrajda and G. Martinelli and C. Pittori and G.C. Rossi}
}

@misc{sandvik_2025,
      title={Using operator covariance to disentangle scaling dimensions in lattice models}, 
      author={Anders W. Sandvik},
      year={2025},
      journal = {arXiv:2406.12681 [cond-mat.stat-mech]},      
      url={https://doi.org/10.48550/arXiv.2406.12681}
}

@article{efe_2023,
    title="{Compression theory for inhomogeneous systems}", 
    author={Doruk Efe Gökmen and Sounak Biswas and Sebastian D. Huber and Zohar Ringel and Felix Flicker and Maciej Koch-Janusz},
    year={2023},
    journal = {arXiv:2301.11934 [cond-mat.stat-mech]},
    url={https://doi.org/10.48550/arXiv.2301.11934}
}

@article{efe_2021,
	title = {Symmetries and phase diagrams with real-space mutual information neural estimation},
	author = {G\"okmen, Doruk Efe and Ringel, Zohar and Huber, Sebastian D. and Koch-Janusz, Maciej},
	journal = {Phys. Rev. E},
	volume = {104},
	issue = {6},
	pages = {064106},
	numpages = {17},
	year = {2021},
	publisher = {American Physical Society},
	url = {https://link.aps.org/doi/10.1103/PhysRevE.104.064106}
}

@article{efe2_2021,
	title = "{Statistical Physics through the Lens of Real-Space Mutual Information}",
	author = {G\"okmen, Doruk Efe and Ringel, Zohar and Huber, Sebastian D. and Koch-Janusz, Maciej},
	journal = {Phys. Rev. Lett.},
	volume = {127},
	issue = {24},
	pages = {240603},
	numpages = {7},
	year = {2021},
	publisher = {American Physical Society},
	url= {https://link.aps.org/doi/10.1103/PhysRevLett.127.240603}
}

@ARTICLE{ramos_2016,
  title    = "Symanzik improvement of the gradient flow in lattice gauge
              theories",
  author   = "Ramos, Alberto and Sint, Stefan",
  journal  = "The European Physical Journal C",
  volume   =  76,
  number   =  1,
  pages    = "15",
  month    =  jan,
  year     =  2016
}

@article{hasenfratz_1993,
title = {Perfect lattice action for asymptotically free theories},
journal = {Nuclear Physics B},
volume = {414},
number = {3},
pages = {785-814},
year = {1994},
issn = {0550-3213},
doi = {https://doi.org/10.1016/0550-3213(94)90261-5},
url = {https://www.sciencedirect.com/science/article/pii/0550321394902615},
author = {P. Hasenfratz and F. Niedermayer}
}

\clearpage

\section{Appendix A - Details of the RSMI-NE procedure}\label{sec:rsmi_details}

For each model, we generate Monte Carlo snapshots at criticality for a fixed linear size $L$. From every snapshot we extract a single training example $(V,E)$. The visible region $V$ is a rasterized disk of radius $r$ centered at a chosen point $c$,
\[
V \equiv \{\,s_x : \lVert x-c\rVert_2 \le r+\tfrac{1}{2}\,\},
\]
and the environment $E$ is a one-lattice-site-thick circular shell at radius $r+b$ around the same center,
\[
E \equiv \{\,s_x : r+b-\tfrac{1}{2} < \lVert x-c\rVert_2 \le r+b+\tfrac{1}{2}\,\}.
\]

The region $V$ is passed through a fully-connected neural network $\Phi_\Theta:\mathbb{R}^{|V|}\to\mathbb{R}$ with parameters $\Theta$ and a single scalar output. We then impose the desired microscopic symmetry sector by projecting $\Phi_\Theta$ onto a one-dimensional irrep of the relevant symmetry group $G$. For $G$ acting on $V$ as $\{P_\pi\}_{\pi\in G}$ and a one-dimensional unitary character $\chi:G\to U(1)$ labelling the target sector, the projection of an arbitrary scalar function $F$ on $V$ is
\begin{equation}
\label{eq:sym_projection}
\mathrm{Sym}_G^{\chi}[F](V) \;\equiv\; \frac{1}{|G|}\sum_{\pi\in G}\chi^\ast(\pi) \, F(P_\pi V).
\end{equation}
The concrete choices of $G$ and $\chi$ for the models studied in this paper are spelled out in Appendix~\ref{sec:sym_sectors}.

Training is performed by maximizing the mutual information between a coarse-grained variable $H$ derived from $V$ and the environment $E$. Concretely, the symmetry-projected scalar $\mathcal{O}_{\mathrm{lat}}(V)$ is first passed through a one-dimensional BatchNormalization layer with learnable shift $\beta$ and scale $\gamma$, where $\gamma$ is hard-constrained by $|\gamma|\le \gamma_{\max}$ (a MaxNorm projection applied after every gradient step). The resulting bounded logit is then fed into a Relaxed-Bernoulli unit at temperature $\tau$,
\[
H \sim \mathrm{RelaxedBernoulli}\!\left(\tau,\, \mathrm{BatchNorm}_{\gamma,\beta}\!\left(\mathcal{O}_{\mathrm{lat}}(V)\right)\right),
\]
with $\tau$ annealed during training. The BatchNorm with bounded $\gamma$ is what realizes the information bottleneck: since the post-BatchNorm signal has variance at most $\gamma_{\max}^2$, the Relaxed-Bernoulli operates in a regime where its sampling noise is non-negligible relative to the logit. Without this constraint, the network could push the logit to arbitrarily large magnitudes and collapse the Relaxed-Bernoulli to a deterministic sign function, pushing the information bottleneck out of the critical regime \cite{gordon_2021}.

An InfoNCE critic network $f_\phi$, with its own parameters $\phi=(\phi_1,\phi_2)$, takes $H$ and $E$ as inputs and provides a variational lower bound $\widehat I_{\mathrm{NCE}}$ on $I(H;E)$. In practice we use a separable critic of the form
\[
f_\phi(H,E) = u_{\phi_1}(H)^{\!\top} v_{\phi_2}(E),
\]
where $u_{\phi_1}$ embeds the scalar bottleneck variable $H$ and $v_{\phi_2}$ embeds the spin variables in the environment shell $E$. The loss $-\widehat I_{\mathrm{NCE}}$ is then minimized by a single Adam step that \emph{jointly} updates the fully-connected network parameters $\Theta$, the BatchNorm parameters $(\gamma,\beta)$, and the critic parameters $\phi$, so that $\Phi_\Theta$ is shaped to maximize the same variational bound that $f_\phi$ certifies. After optimization, the BatchNorm \footnote{ affine recalibrations from BatchNorm do not affect connected correlators.}, the binary head, and the critic are all discarded, and the trained, symmetry-projected neural operator $\mathcal{O}_{\mathrm{lat}} = \mathrm{Sym}_G^\chi[\Phi_\Theta]$ is retained as the lattice operator used throughout the main text.

The full training procedure is summarized in Algorithm~\ref{alg:rsmi_ne}, and the implementation hyperparameters used in this work are listed in Table~\ref{tab:rsmi_hyperparams}.

\begin{table}[h]
\centering
\begin{tabular}{l|c}
\hline
Quantity & Value \\
\hline
System size in RSMI training, $L$ & $40$ \\
Visible region radius, $r$ & $3$ \\
Buffer width, $b$ & $4$ \\
Number of training iterations, $T$ & $10^4$ \\
Batch size, $N$ & $8{,}000$ \\
Learning rate, $\eta$ & $3\times 10^{-4}$ \\
Initial temperature, $\tau_0$ & $1.0$ \\
Final temperature, $\tau_{\min}$ & $0.2$ \\
Temperature decay rate, $\alpha$ & $1.5\times 10^{-4}$ \\
BatchNorm scale bound, $\gamma_{\max}$ & $0.1$ \\
Architecture of $\Phi_\Theta$ & 3 FC$\times 2\left|V\right|$, ReLU $\to 1$ \\
Architecture of $u_{\phi_1}$ & 3 FC$\times 64$, ReLU $\to 256$ \\
Architecture of $v_{\phi_2}$ & 3 FC$\times 2\left|E\right|$, ReLU $\to 256$ \\
\hline
\end{tabular}
\caption{RSMI-NE training hyperparameters used to construct the neural lattice operators, common to all three models studied. ``$3$ FC$\times w$, ReLU $\to d$'' denotes three fully-connected hidden layers of width $w$ with ReLU activations, followed by a linear projection to dimension $d$. The temperature decay rate satisfies $\alpha = \log(\tau_0/\tau_{\min})/(0.95\,T)$, so that $\tau_t$ reaches $\tau_{\min}$ after $95\%$ of training.}
\label{tab:rsmi_hyperparams}
\end{table}

\begin{figure*}[t]
\centering
\begin{minipage}{0.96\textwidth}

\hrule
\vspace{0.6ex}
\noindent\textbf{RSMI-NE training of an improved lattice operator}
\label{alg:rsmi_ne}
\vspace{0.6ex}
\hrule

\vspace{1.0ex}
\label{alg:rsmi_ne}

\vspace{0.5em}
\label{alg:rsmi_ne}
\begin{algorithmic}[1]

\Statex \textbf{Goal.} Given a dataset of visible/environment pairs $\{(V,E)\}$ extracted from Monte Carlo configurations of a 2D spin system at criticality, learn a local neural operator $\mathcal{O}_{\mathrm{lat}}(V)$ with large overlap with the leading long-distance operator in a chosen symmetry sector.

\Statex
\Statex \textbf{Training loop.}
\For{$t=1,\dots,T$ \Comment{$T$: number of training iterations}}
   \State $\tau_t \gets \max\!\left(\tau_{\min},\, \tau_0 e^{-\alpha t}\right)$ \Comment{anneal the binary-head temperature}
   \State Draw a batch $\{(V^{(i)},E^{(i)})\}_{i=1}^N$ \Comment{$N$: batch size}

   \State Forward pass through the symmetrized network, BatchNorm, and binary head:
   \Statex \hspace{2.2em}
   $\ell^{(i)} \gets \mathrm{Sym}_G^{\chi}\!\left[\Phi_\Theta\right]\!\left(V^{(i)}\right)$
   \hfill (symmetric smooth scalar)
   \Statex \hspace{2.2em}
   $\tilde\ell^{(i)} \gets \mathrm{BatchNorm}_{\gamma,\beta}\!\left(\ell^{(i)}\right)$
   \hfill ($|\gamma|\le\gamma_{\max}$; bounded logit / information bottleneck)
   \Statex \hspace{2.2em}
   $H^{(i)} \sim \mathrm{RelaxedBernoulli}\!\left(\tau_t,\, \tilde\ell^{(i)}\right)$
   \hfill (soft binary surrogate)

   \State Maximize the InfoNCE lower bound on $I(H;E)$:
   \Statex \hspace{2.2em}
   $s_{ij} \gets f_\phi\!\left(H^{(i)},E^{(j)}\right)$
   \hfill ($N\times N$ critic-score matrix)
   \Statex \hspace{2.2em}
   \[
   \widehat I_{\mathrm{NCE}}
   \gets
   \frac{1}{N}
   \sum_{i=1}^N
   \log
   \frac{\exp s_{ii}}
   {\frac{1}{N}\sum_{j=1}^N \exp s_{ij}} .
   \]
   \Statex \hspace{2.2em}
   $(\Theta,\gamma,\beta,\phi) \gets (\Theta,\gamma,\beta,\phi)+\eta\,\nabla\widehat I_{\mathrm{NCE}}$,\quad then project $\gamma\to\mathrm{clip}(\gamma,\,[-\gamma_{\max},\gamma_{\max}])$
   \hfill (Adam step; $\eta$: learning rate)
\EndFor

\Statex
\Statex \textbf{Output.} \Return the neural operator
\[
\mathcal{O}_{\mathrm{lat}}(V)
=
\mathrm{Sym}_G^{\chi}\!\left[\Phi_\Theta\right]\!(V).
\]

\end{algorithmic}
\vspace{1.0ex}
\hrule
\end{minipage}
\end{figure*}

\section{Appendix B -- Symmetry sectors for the Ising and Potts models}\label{sec:sym_sectors}

In both models discussed below, the lattice symmetry group factorizes as $G = G_{\mathrm{int}}\times D_4$, where $G_{\mathrm{int}}$ is the on-site internal symmetry and $D_4$ is the lattice point group of $90^\circ$ rotations and mirror reflections. All operators we target are scalars under the point group, so we always take the trivial $D_4$ character, $\chi(\pi)=+1$ for every $\pi\in D_4$; the symmetry sector is therefore determined entirely by the one-dimensional character of $G_{\mathrm{int}}$.

For the Ising model, $G_{\mathrm{int}}=\mathbb{Z}_2$ (global spin flip $s_x\mapsto -s_x$) and the two real characters $\chi\in\{\pm 1\}$ correspond to the two leading scalar primaries: the energy operator $\epsilon$ (trivial $\mathbb{Z}_2$ character, $\chi=+1$) and the spin operator $\sigma$ ($\mathbb{Z}_2$-odd character, $\chi=-1$).

For the $q=3$ Potts model, $G_{\mathrm{int}}=\mathbb{Z}_3$ and the three one-dimensional characters are $\{1,\omega,\omega^2\}$ with $\omega=e^{2\pi i/3}$. Let $F_k(V)$ denote the $D_4$-symmetrized output of $\Phi_\Theta$ after the global Potts shift $s_x\mapsto s_x+k\bmod 3$ (with $k=0,1,2$). The $\mathbb{Z}_3$-invariant character $\chi\equiv 1$ selects the $\epsilon$ sector, whose lattice operator is the symmetric sum
\[
\mathcal O_\epsilon(V)
=
\frac{1}{3}\left(F_0(V)+F_1(V)+F_2(V)\right).
\]
The two complex-conjugate characters $\chi\in\{\omega,\omega^2\}$ select the $\sigma$ sector; to avoid complex-valued operators, we work in their real cosine/sine basis,
\[
\mathcal O_\sigma^{\cos}(V)
=
\frac{1}{3}\left(F_0(V)-\frac{1}{2}F_1(V)-\frac{1}{2}F_2(V)\right),
\]
\[
\mathcal O_\sigma^{\sin}(V)
=
\frac{1}{\sqrt{2}}\left(F_1(V)-F_2(V)\right),
\]
which together span the leading Potts spin sector.

\section{Appendix C -- Comparison of neural and naive correlation functions}\label{sec:correlation_comparison}

\begin{figure}[htb]
    \centering
    \includegraphics[width=0.4\textwidth]{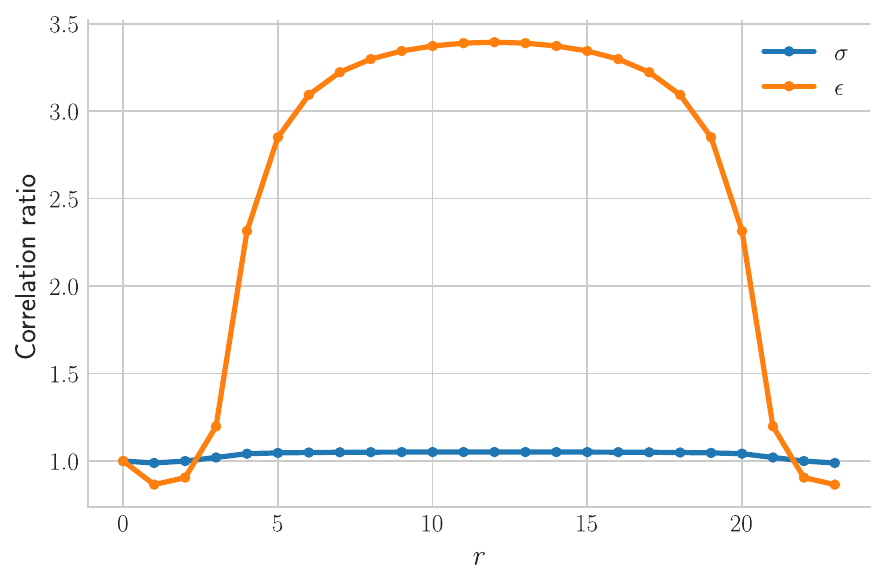}
    \caption{Ratio $R_O(r)$ of connected two-point functions for the neural and naive operators in the Ising model at $L=24$. The enhancement is substantial for the energy operator $\epsilon$ and modest for the spin operator $\sigma$, consistent with the larger reduction of finite-size corrections observed for $\epsilon$ in the main text.}
    \label{fig:correlation}
\end{figure}

As an auxiliary diagnostic, we compare the connected two-point functions of the learned operators with those of the conventional lattice estimators at fixed system size. For a given operator $O$, we consider the ratio
\begin{equation}
R_O(r)=
\frac{\langle O_{\mathrm{neural}}(0)\,O_{\mathrm{neural}}(r)\rangle/Var\left(O_{\mathrm{neural}}\right)}
     {\langle O_{\mathrm{naive}}(0)\,O_{\mathrm{naive}}(r)\rangle/Var\left(O_{\mathrm{naive}}\right)}
\end{equation}
Figure~\ref{fig:correlation} shows this quantity for the Ising model at $L=24$.

For the energy operator $\epsilon$, the ratio develops a broad plateau at intermediate and long distances with a value of approximately $3.5$, indicating that the learned estimator produces a substantially stronger long-distance signal than the naive lattice operator. By contrast, for the spin operator $\sigma$, the corresponding ratio remains close to unity, around $1.1$, showing only a modest enhancement. The same qualitative pattern is observed in the $q=3$ Potts and dilute $q=3$ Potts model.

\end{document}